\documentclass[]{spie}  

\usepackage{amsmath,amsfonts,amssymb}
\usepackage{graphicx}
\usepackage[colorlinks=true, allcolors=blue]{hyperref}

\title{Design study and first performance simulation of the ELT/MICADO focal plane coronagraphs}

\author[a]{Cl\'ement Perrot}
\author[a]{Pierre Baudoz}
\author[a]{Anthony Boccaletti}
\author[a]{G\'erard Rousset}
\author[a]{Elsa Huby}
\author[a]{Yann Cl\'enet}
\author[a]{S\'ebastien Durand}
\author[b]{Richard Davies}
\affil[a]{LESIA, Observatoire de Paris, PSL Research University, CNRS, Sorbonne Universit\'es, UPMC Univ. Paris 06, Univ. Paris Diderot, Sorbonne Paris Cit\'e, 5 place Jules Janssen, 92190 Meudon, FRANCE}
\affil[b]{Max-Planck-Institut fur extraterrestrische Physik, Garching, GERMANY}

\authorinfo{Further author information: (Send correspondence to Perrot Cl\'ement.)\\Perrot Cl\'ement: E-mail: clement.perrot@obspm.fr\\  Baudoz Pierre : E-mail: pierre.baudoz@obspm.fr}

\begin{document} 
\maketitle

\begin{abstract}

In this paper, we present the design and the expected performance of the classical Lyot coronagraph for the high contrast imaging modes of the wide-field imager MICADO. MICADO is a near-IR camera for the Extremely Large Telescope (ELT, previously E-ELT), with wide-field, spectroscopic and coronagraphic capabilities. MICADO is one of the first-light instruments selected by the ESO. Optimized to work with a multi-conjugate adaptive optics corrections provided by the MOARY module, it will also come with a SCAO correction with a high-level, on-axis correction, making use of the M4 adaptive mirror of the telescope.

After presenting the context of the high contrast imaging modes in MICADO, we describe the selection process for the focal plane masks and Lyot stop. 
We will also show results obtained in realistic conditions, taking into account AO residuals, atmospheric refraction, noise sources and simulating observations in angular differential imaging (ADI) mode. Based on SPHERE on-sky results, we will discuss the achievable gain in contrast and angular separation provided by MICADO over the current instruments on 10-m class telescopes, in particular for imaging young giant planets at very short separations around nearby stars as well as planets on wider orbits around more distant stars in young stellar associations.

\end{abstract}


\section{INTRODUCTION}
\label{sec:intro}  

MICADO \cite{2016SPIE.9908E..1ZD} is the first-light near-infrared camera on the ELT, which will work between 0.8\,$\mu$m and 2.5\,$\mu$m. MICADO is planned to work at the diffraction limit over the whole field-of-view of the instrument, thanks to the multi-conjugate adaptive optics (MCAO) correction provided by the MAORY AO module \cite{2016SPIE.9909E..2DD}. To complement this MCAO correction, a single conjugate adaptive optics (SCAO) mode is developed under MICADO's management, jointly with MAORY, and will be the first commissioned AO mode  at the ELT \cite{2016SPIE.9909E..0AC}.

Exoplanet detection and characterization is one of the main science cases of the ELT.  Pending the installation of a dedicated facility on the ELT to search and characterize mature exoplanets (like ELT/PCS\cite{2013aoel.confE...8K}),  the SCAO mode of MICADO-MAORY will be able to address this prominent goal at the ELT first light, in particular thanks to the increase of angular resolution by a factor of 5 with respect to the VLT. MICADO come online when planet finders on 8-m class telescopes (SPHERE/VLT, GPI/Gemini) will have provided their main results (large discovery surveys) and after the start of JWST operations. The exoplanets science case described below fits in this general context.

Dues to the constraints of direct imaging (the angular separation and the contrast between the star and the planet), past and current surveys focus on young stars close to the Sun ($<$\,150\,pc). Observing the very first steps of the planetary formation with direct imaging brings important information about the formation process. Direct imaging provides not only the means to witness newly formed planets, but it also provides a clear view of the environment of the star, in particular during the protoplanetary phase when planets are in formation. Imaging the circumstellar disk gives a lot of information about the evolution of the systems and its composition. In some particular cases the planet is embedded in the circumstellar disk, like in the beta Pictoris system, where a 10 Jupiter mass ($M_J$) companion was discovered \cite{2009A&A...493L..21L} within the debris disk \cite{1984Sci...226.1421S} which presents a warp caused by the planet.

Direct imaging so far has only revealed a handful of planetary mass objects due to the high star to planet contrast. 
During the past few years, instruments dedicated to high contrast imaging, like Gemini/GPI \cite{2008SPIE.7015E..18M} and VLT/SPHERE \cite{2008SPIE.7014E..18B}, have come online on 10-m class telescope and provide deep observations close to the star thanks to their extrem AO system. These instruments have led to new detections of objects like the two exoplanets 51 Eridani b \cite{2015Sci...350...64M} and HIP 65426 \cite{2017A&A...605L...9C} but have also improved the charaterization of known exoplanets atmosphere like HR 8799 \cite{2016A&A...587A..57Z} \cite{2016A&A...587A..58B}. Circumstellar disk detection was also a prolific area with for instance the debris disk HD 106906 \cite{2016A&A...586L...8L} or the transition disk SAO 206462 \cite{2017A&A...601A.134M}.

In this context, MICADO will provide deep observation close to the star with respect to GPI and SPHERE without the need for an extreme adaptive optics system. The very-high angular resolution offered by MICADO (8.7 mas in H-band) combined with the AO correction will be particularly suited to explore the range of orbite between 40 and 150\,mas, that is inaccessible for 10-m telescope.

\section{The coronagraphic mode of MICADO}

The current design of MICADO allows the implementation of coronagraphic modes necessary to achieve the science cases described above. However, as MICADO is not dedicated to high contrast imaging like VLT/SPHERE, some trade-off are inevitable. The science program that can be accomplished strongly depends on the achieved Strehl ratio. Since this program will observe bright on-axis stars, a SCAO correction is mandatory to reach the highest possible Strehl ratio.

\subsection{Coronagraphic constraint}

Because of its large central obscuration and its segmented pupil, the ELT pupil is not optimized for coronagraphy. One solution to overcame this problem is to use an apodisation mask at the entrance pupil of the instrument to compensate the effect of the diffraction by the central obscuration. However, the final design of MICADO does not permit it. In this context, a simple Classical Lyot Coronagraph (CLC), without apodisation was selected in the final design. The CLC is composed by an occulting focal plane mask and a Lyot stop placed in a pupil plane downstream of the focal plane mask. All these components are located in the cryostat. Moreover, the Atmospheric Dispersion Compensator (ADC) will be placed between the entrance focal plane and the intermediate pupil plane. In this condition, focal plane masks will be subjected to atmospheric dispersion.

In the current design of MICADO, the coronagraphic mode has 3 slots available for focal plane masks in a wheel at the entrance focal plane of the cryostat and 5 slots for pupil masks in the wheel at the intermediate pupil plane. In the focal plane, 2 of these slots are reserved for the CLC and the last one will be a vortex phase mask. In the pupil plane, 1 slot is reserved for the Lyot stop, 2 slots for aperture masking \cite{2014SPIE.9147E..9FL} and 2 slots for vAPP coronagraph \cite{2012SPIE.8450E..0MS}. This study focuses on the CLC definition. For that purpose, our simulations are based on the SCAO simulations performed with COMPASS \cite{2014SPIE.9148E..6OG}, which give a Strehl ratio of 72\% at 2.2$\mu m$.

\subsection{Others requirements}

The high contrast imaging mode also requires some specifics component, in particular in the double wheel filters. For the good operation of the coronagraphic mode, it is important to observe with some special filter and neutral density at the same time. Consequently, all scientific filters for the exoplanetary sciences are necessarily placed in the first wheel, whereas, neutral density sit in the second wheel. Table \ref{tabfiltre} shows the requirements of the filters for exoplanets sciences in the current design.
High contrast imaging also requires the ability to observe in pupil tracking (PT) mode, which corresponds to an observation mode which keeps the pupil stabilized with respect to the instrument while the field rotates on the detector. For that purpose, the derotator is operated at a different speed than in the field tracking mode. The PT mode is mandatory to apply post-processing algorithms based on Angular Differential Imaging (ADI) that are necessary to achieve deep detection limits.

\begin{table}[!h]
\centering
\begin{tabular}{ l c c c c}
\hline
Filter  & $\lambda_{c}$  & $\Delta\lambda$  & $\lambda_{min}$  & $\lambda_{max}$ \\ 
- & [$\mu m$] & [nm] & [$\mu m$] & [$\mu m$] \\ 
\hline 
J & 1,245 & 180 & 1,155 & 1,335 \\ 
H & 1,635 & 290 & 1,490 & 1,780 \\ 
Ks & 2,145 & 350 & 1,970 & 2,320 \\ 
J1-short & 1,190 & 50 & 1,165 & 1,215 \\ 
J2-short & 1,270 & 50 & 1,245 & 1,295 \\ 
CH4-short & 1,582 & 85 & 1,540 & 1,625 \\ 
CH4-long & 1,693 & 112 & 1,637 & 1,749 \\ 
H20\_K & 2,060 & 60 & 2,030 & 2,090 \\ 
COabs & 2,308 & 44 & 2,286 & 2,330 \\ 
K1-mid & 2,100 & 100 & 2,050 & 2,150 \\ 
K1-cont & 2,202 & 29 & 2,188 & 2,216 \\ 
\hline
\end{tabular}
\caption{List of MICADO's filters for exoplanetary sciences.}
\label{tabfiltre}
\end{table}			

\section{Simulation tool}

To select the design of the CLC and estimate the performance of the coronagraphic mode of MICADO, we used an aberration phase map after correction by the SCAO. The GPU-based tool COMPASS, developed at Paris Observatory, delivers these AO corrected phase maps. These AO corrected phase maps feed a dedicated code which simulates a coronagraphic observation including the atmospheric dispersion, polychromatisme, field rotation, etc \cite{2014SPIE.9147E..9EB}.

\subsection{COMPASS}
\label{sec:compass}

COMPASS is a tool which can simulate various AO systems and observations. For this study we used the following hypotheses in the AO simulation:
\begin{itemize}
\item[-] 39.146\,m circular pupil with central obscuration of 24\%, no spiders, no segments,
\item[-] Shack-Hartmann with 78x78 sub-apertures, working at 500\,Hz,
\item[-] $r_0$ = 0.129\,m at 500\,nm, wind = 10\,m.s$^{-1}$, $L_0$ = $10^5$
\item[-] One frame is extracted every 500 simulated residual phase screens to create a cube of 3600 AO corrected phase maps (1 map per second, for a total of 1 hour observation).
\end{itemize}
Note that the baseline design for the MICADO-MAORY SCAO relies one a Pyramid wave front sensor (WFS). However, this kind of sensor was not yet fully debugged at the time of the present work. Hence a Shack-Hartmann WFS has been used providing similar results as the Pyramid WFS for the considered level of correction (Strehl ratio at 2.2\,$\mu$m = 70\%).

\begin{figure}[!h]
\centering
\includegraphics[width=0.24\textwidth,clip]{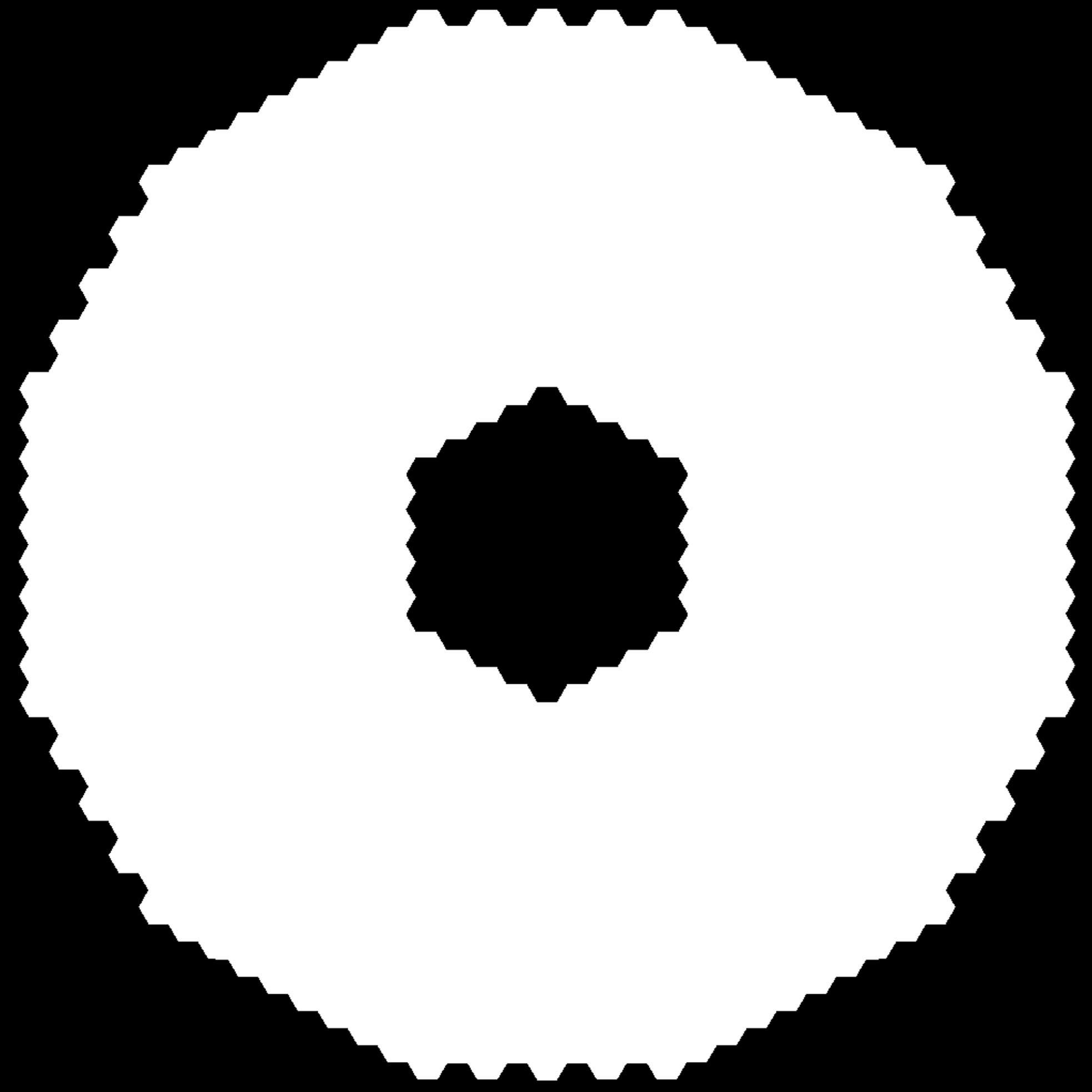}
\includegraphics[width=0.24\textwidth,clip]{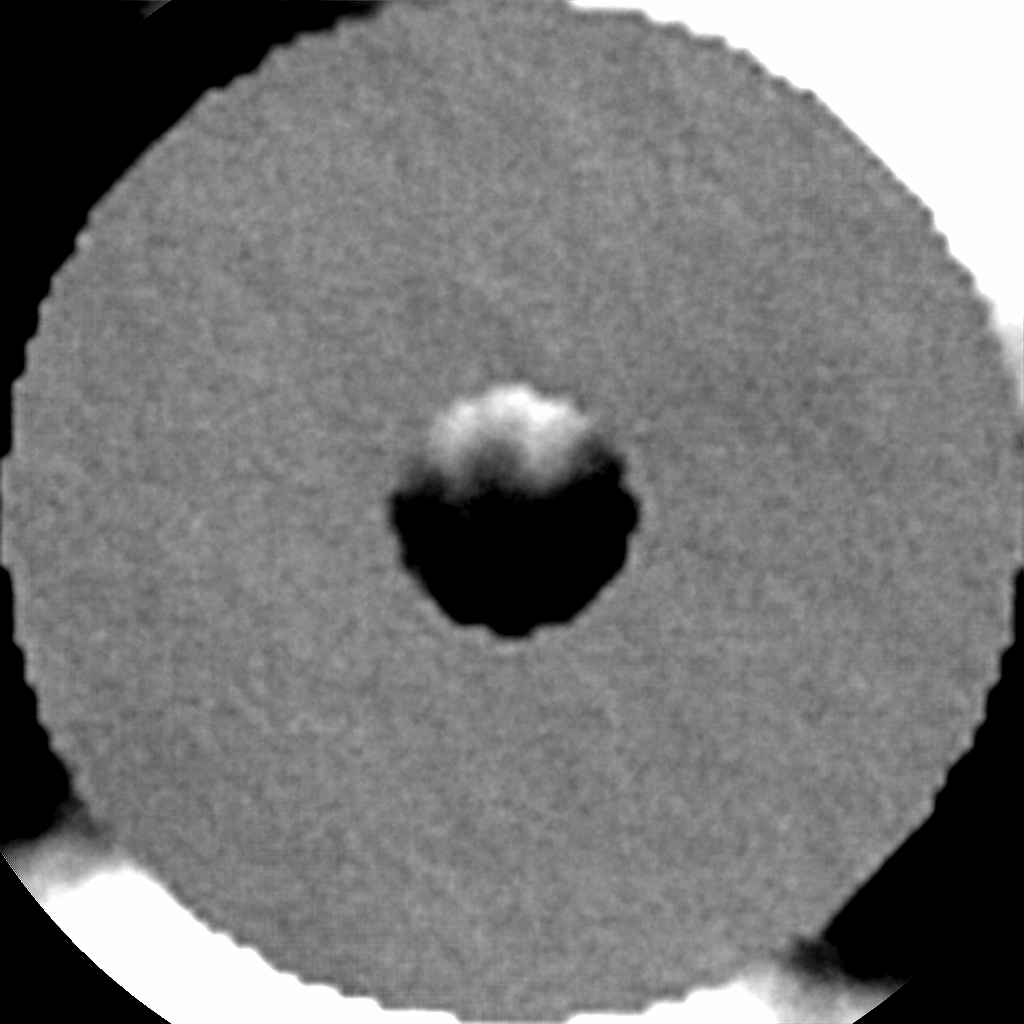}
\includegraphics[width=0.24\textwidth,clip]{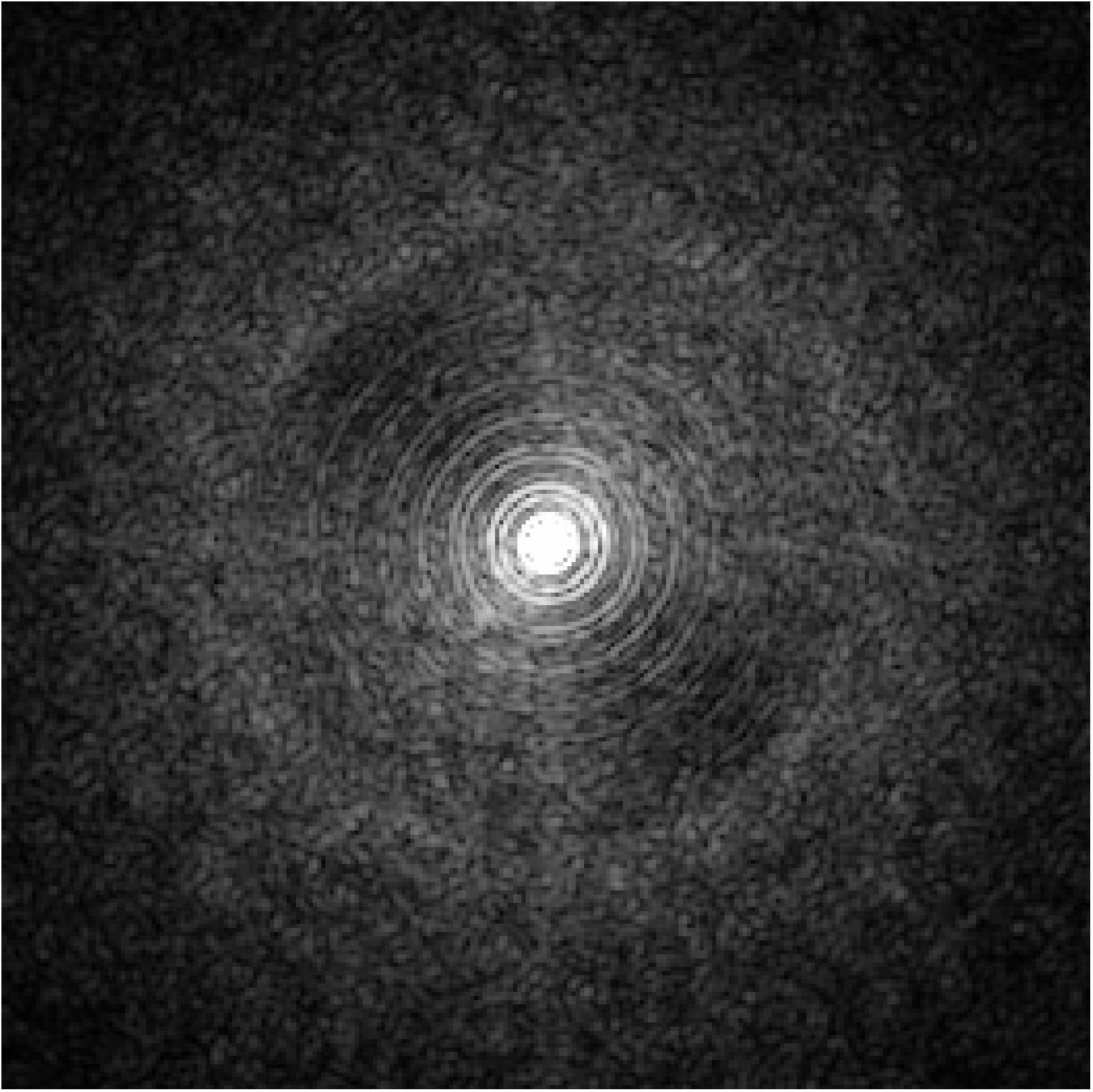}
\includegraphics[width=0.24\textwidth,clip]{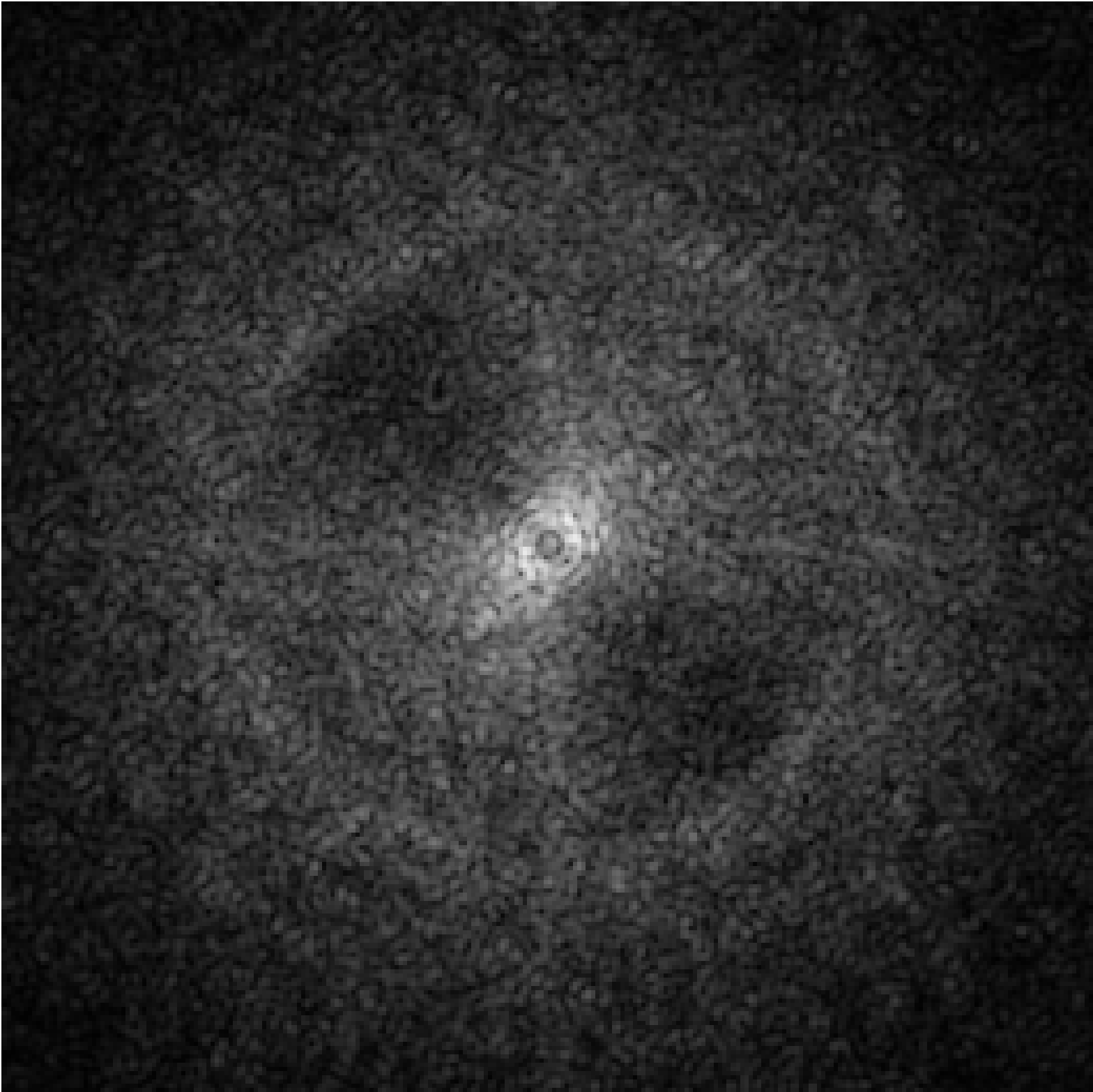}
\caption{From left to right: the ELT pupil, a typical residual phase map from COMPASS, a typical instantaneous PSF and a typical instantaneous coronagraphic image.}
\label{structMICADO}
\end{figure}

\subsection{Coronagraph simulation}
\label{sec:corono}

The 3600 AO-corrected phase maps calculated with COMPASS are sent to an IDL-based simulation tool dedicated to the estimation of high contrast performance of MICADO \cite{2014SPIE.9147E..9EB}. This simulation tool can simulate different types of coronagraph to the ELT pupil, add a set of static and/or quasi-static additional aberrations, do polychromatic simulation, include the atmospheric dispersion on each wavelength and perform the field rotation as a function of the declination of the target.
The typical correlation time of quasi-static speckles and the amplitude of static and quasi-static speckles are difficult to estimate at this early stage of the telescope and instrument development. However, on a Nasmyth focus where MICADO will be installed, PT mode requires the rotation of optical elements. Thus we decided to introduce a fixed aberration phase map with respect to the instrument and an aberration phase map which rotates with the field of view. For the amplitude of the static and quasi-static aberrations, we based our estimation on the SPHERE estimation and we increased the amplitude. The specific hypotheses for the coronagraph simulation are given below:
\begin{itemize}
\item[-] The CLC is defined by 3 parameters: $R_M$ the radius of the focal plane mask, $D_e$ the outer diameter of the Lyot stop and $D_i$ the inner diameter of the Lyot stop. $D_e$ and $D_i$ are defined in fraction of the pupil diameter.
\item[-] In case of polychromatic simulation: for each wavelength, the angular shift due to the atmospheric dispersion is computed and applied in the coronagraphic mask plane, following the model of \cite{1985spas.book.....G}. We used the reference values of atmospheric parameters given by ESO for the ELT : $P_{atmo}$ = 712\,mbar, $T_{atmo}$ = 9,1\,$^{\circ}$C and $H_{atmo}$ = 15\,\%.
\item[-] Static aberrations have amplitude of 60\,nm\,RMS, with a variation of $\pm$6\,nm\,RMS which corresponds to the quasi-static aberrations. These values are derived from the experience with the SPHERE instrument \cite{2008SPIE.7015E..6EB}. The Power Spectral Density (PSD) of the aberrations is chosen to vary as $f^{-2}$ with $f$ the spatial frequency of the defects.
\item[-] The static aberration map is fixed, while the quasi-static aberration map rotates with the field of view. The rotation is given by the parallactic angle of the star, and it is set such that the star crosses the south meridian at the middle of the observation. The parallactic angle of each coronagraphic image is determined by the declination of the star, the total time of integration and the time of integration for a single coronagraphic image.
\item[-] For each wavelength and for each AO corrected phase map, a instantaneous coronagraphic and a non coronagraphic image are created. These images are stacked depending on the individual integration time (ex.: for an image of 10 seconds we stack 10 instantaneous image) and of the spectral bandwidth.
\end{itemize}

\subsection{Coronagraph selection}
\label{sec:selec}

To take full advantage of the high angular resolution of the ELT, we decided to select a CLC with a focal plane mask with a small Inner Working Angle (IWA). For the second CLC we choose a focal plane mask with a medium IWA to take care of the PSF dispersion and to ensure a good sensibility of the coronagraphic mode. The radius of the small focal plane mask is called $R_{M1}$ and the $R_{M2}$ for the medium focal plane mask.
The selection of the size of the 2 focal plane masks is linked to the selection of the dimension of the Lyot stop. The size of the 3 elements is optimized at the same time. Simulations were performed in monochromatic light at 2.2\,$\mu$m with 10 AO corrected phase maps (corresponding to a 10 seconds image). 
To do so, we defined a criterion that optimizes the contrast close to the center of the focale plane mask ($<$\,6$\lambda$/D) and the attenuation of the central peak in the coronagraphic image. This attenuation is mandatory to have enough dynamic range and avoid the saturation of the detector.\\

\begin{figure}[!h]
\centering
\includegraphics[width=0.49\textwidth,clip]{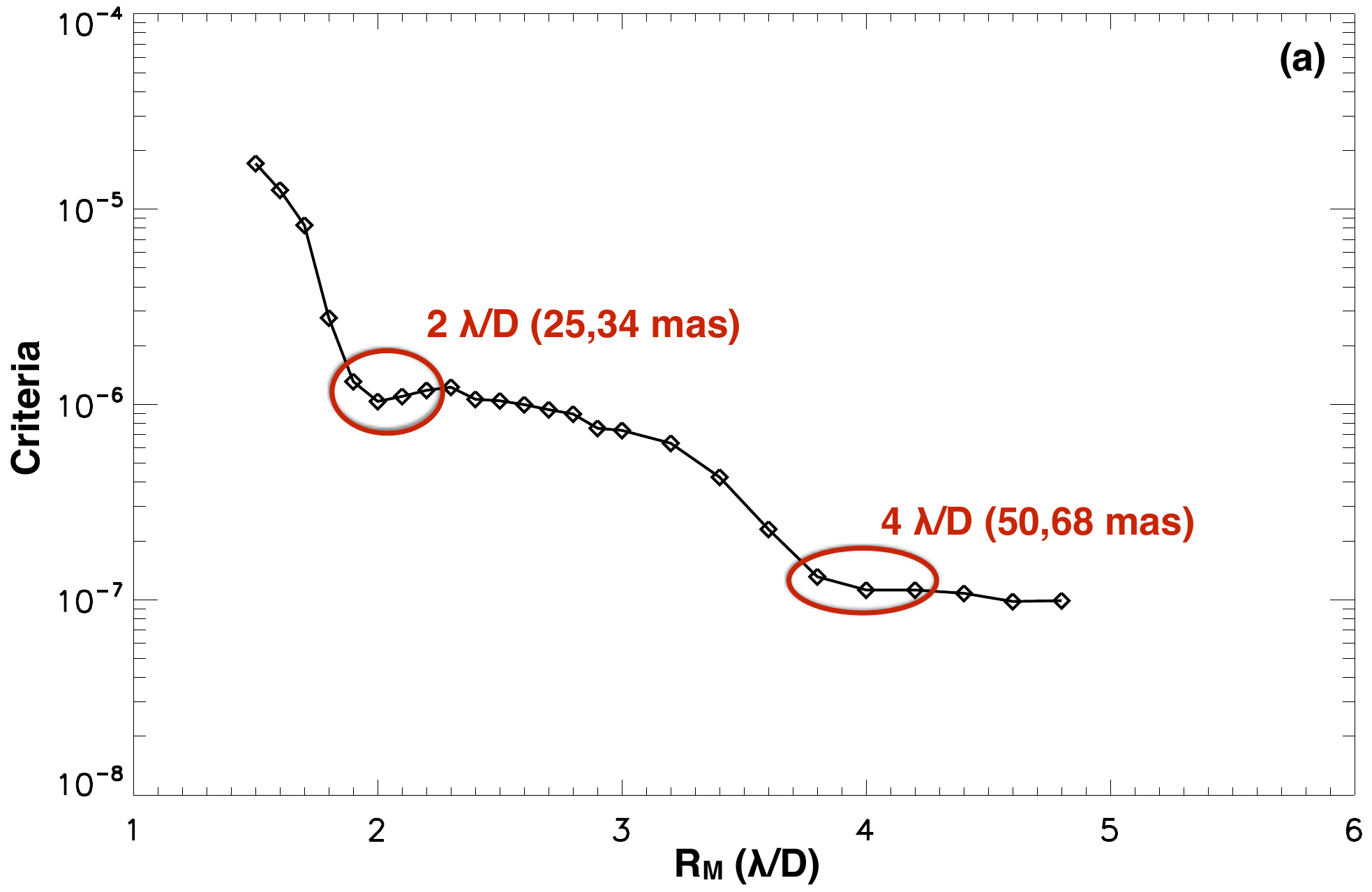}
\includegraphics[width=0.49\textwidth,clip]{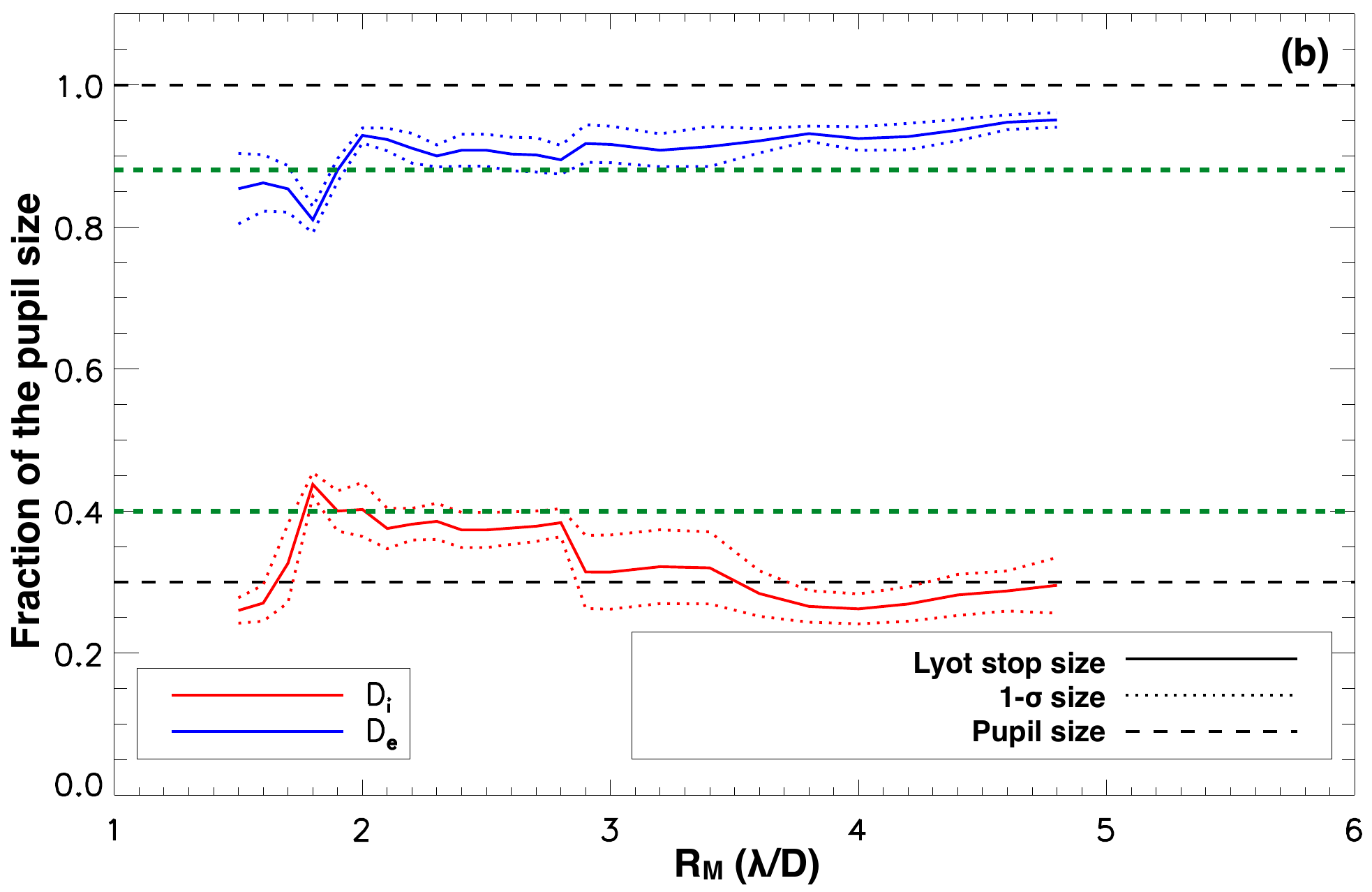}
\caption{Left: the selection criterion for the CLC, which combines best contrast at small angular separation and best attenuation of the star as a function of the radius of the focal plane mask in $\lambda$/D (at 2.2\,$\mu$m). Two stages, hightlighted in the red circles, show the selected masks. Right: the size of the Lyot stop in fraction of the pupil size, as a function of the radius of the focal plane mask. Red lines are for the size of the obscuration of the Lyot stop, blue lines are for the size of the outer part of the Lyot stop and green lines are the selected size of the outer part and the obscuration of the Lyot stop.}
\label{selectionmask}
\end{figure}

We thus simulate coronagraphic images for several sizes of focal plane mask and Lyot stop. For each image we measure the mean contrast between $R_{M1}$+1$\lambda$/D and 6$\lambda$/D and the attenuation, defined as the maximum of the coronagraphic image divided by the maximum of the non-coronagraphic image. The multiplication of these two quantities gives the selection criterion, which is optimal when it is low.
Figure \ref{selectionmask} (a) shows the best values of the criterion as function of the radius of the focal plane mask. We identify 2 sizes of mask which match with the requirement at 2$\lambda$/D and 4$\lambda$/D, for $\lambda$=2.230\,$\mu$m (25.34 mas and 50.68 mas).
The figure \ref{selectionmask} (b) shows the optimal size of the Lyot stop for each size of the focal plane mask. The size is given in fraction of the pupil size, the red line corresponds to the size of the central obscuration and the blue line to the size of the outer part of the Lyot stop. As only one Lyot stop is available for both masks, we decide to select the Lyot stop that optimizes the small focal plane mask. The selected Lyot mask has thus an obscuration of 40\% of the pupil diameter and an outer diameter of 88\% of the pupil diameter. The total transmission of the Lyot stop, with respect to the ELT pupil is 66,1\%.

\section{Effect of atmospheric refraction}
\label{sec:atmo}

\begin{figure}[!h]
\centering
\includegraphics[width=0.54\textwidth,clip]{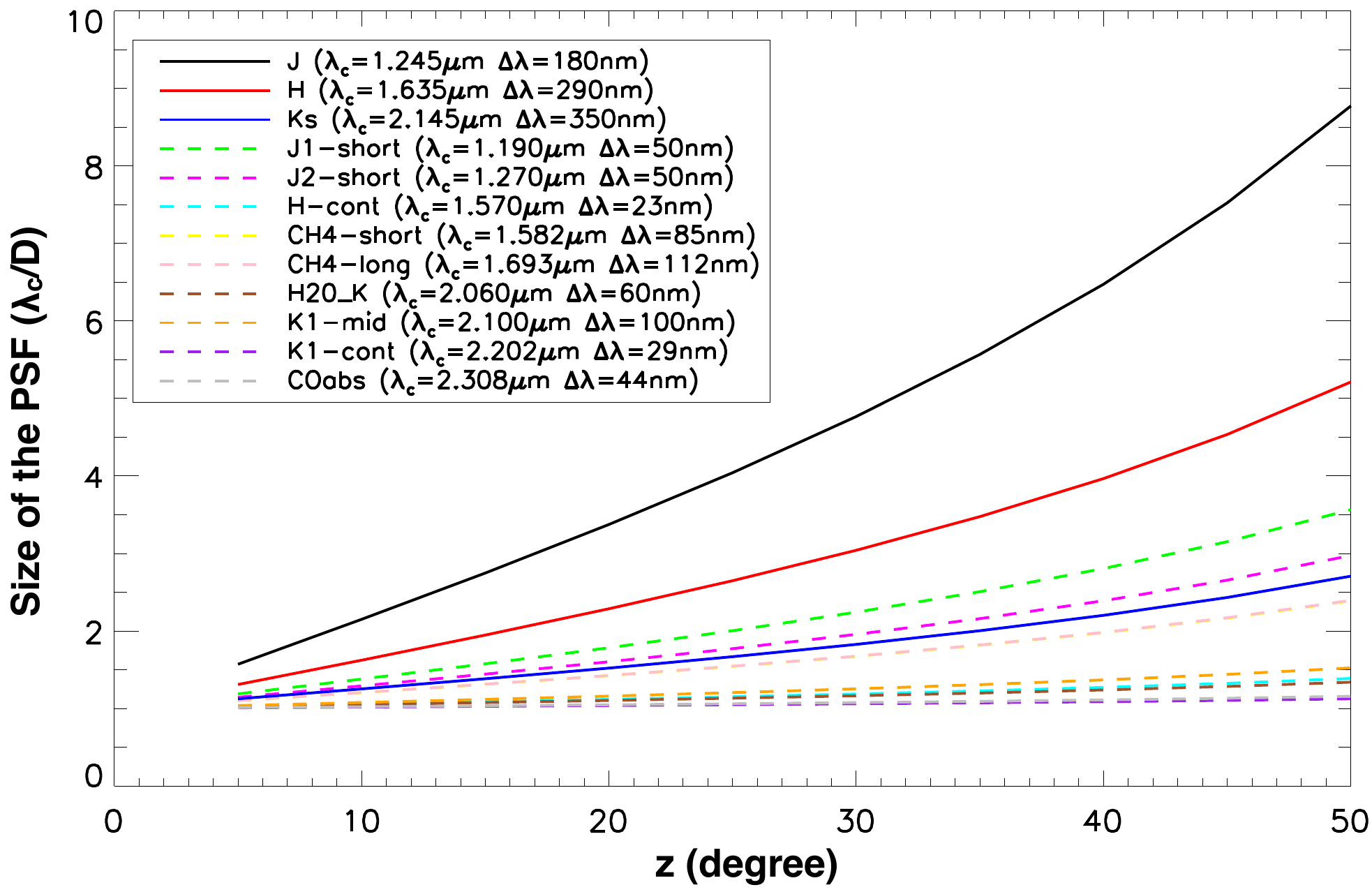}
\includegraphics[width=0.45\textwidth,clip]{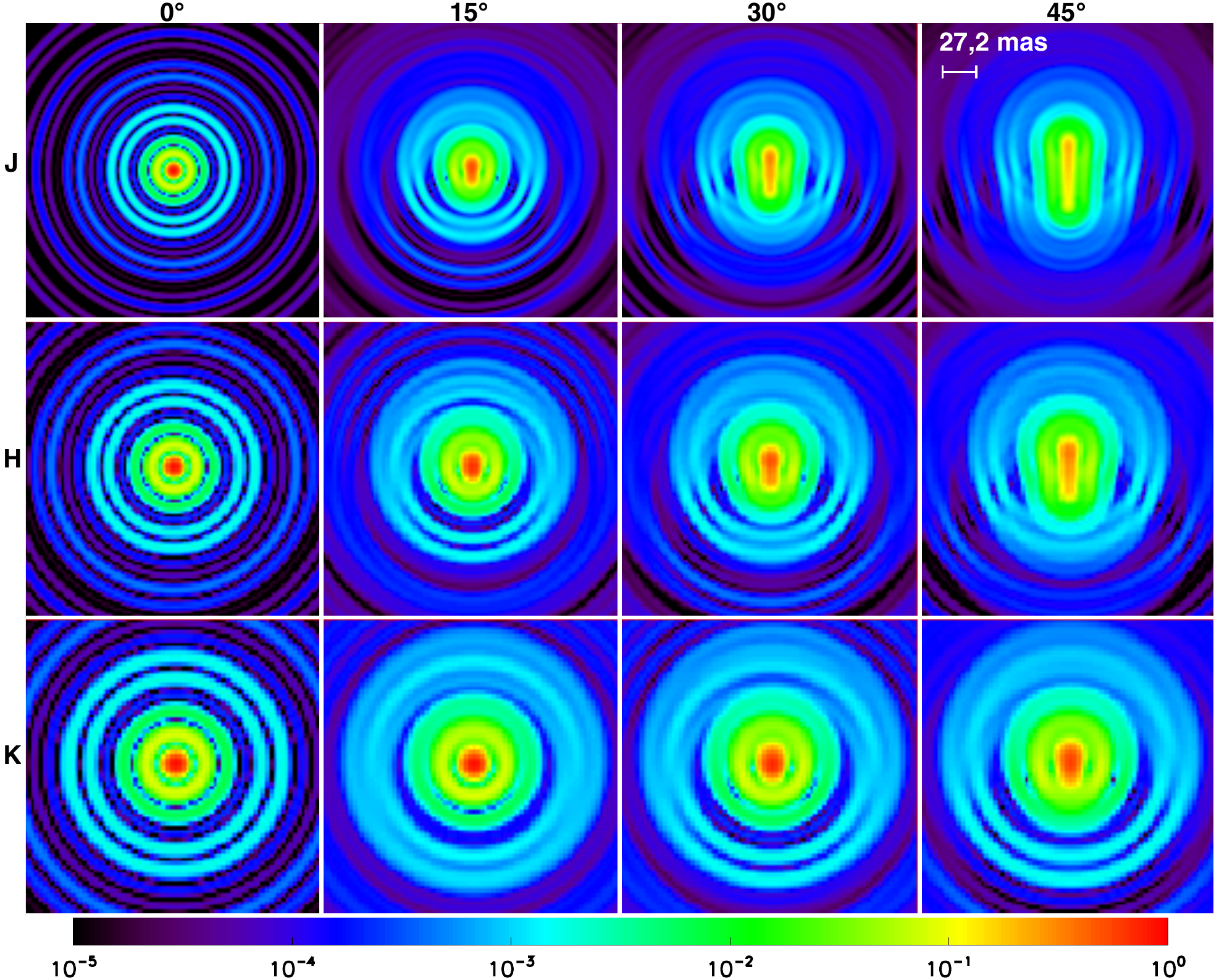}
\caption{Left: size of the PSF, following the atmospheric dispersion, in $\lambda_c$/D as a function of the zenithal angle $z$ and for all MICADO's filters. Right: typical PSF shape deformed by the atmospheric dispersion with broadband filters J, H and K, for several values of the zenithal angle (0, 15, 30 and 45 degrees) in the case of standard atmospheric conditions.}
\label{dispi}
\end{figure}

As explained previously, the ADC is placed after the focal plane where coronagraphic masks are located. This design has an impact on the high contrast imaging mode because the CLC will be less efficient due to the PSF elongation. Using reference values of the atmospheric parameters, we estimate the effect of the dispersion on the coronagraphic image. To simulate the dispersion effect of the atmosphere we used the approximation of \cite{1985spas.book.....G} :
\begin{equation}
n_0(\lambda) \sin{z_0(\lambda)} = \sin{z}.
\end{equation}
Where $n_0(\lambda)$ is the refraction index of the atmosphere at the telescope, $z_0(\lambda)$ the zenithal angle of the star seen by the telescope and $z(\lambda)$ the real zenithal angle of the star.
Figure \ref{dispi} (a) shows the size of the PSF (in the direction of the elongation) as a function of the zenithal angle (z) and figure \ref{dispi} (b) shows some examples of PSF in J, H and K bands with a zenithal angle of 0 (no dispersion), 15, 30 and 45 degrees. Narrowband filters should not be significantly affected by the dispersion, especially in K band. However, broadband filters are more sensitive to the dispersion. The broadband filter Ks is not very affected by the dispersion when le zenithal angle is low, but in case of targets close to the horizon, the dispersion is very strong. Finally the two broadband filters H and J are strongly affected by the dispersion. In these bands, the medium focal plane mask is mandatory, while the small mask is usable for the other filters.

\section{One hour ADI observation simulation}
\label{sec:adi}

To estimate the on-sky performance of the coronagraphic mode of MICADO we performed ADI simulations, that are then post-processed and compared to on-sky observation obtained with SPHERE. For these simulations we assume 360 coronagraphic images, corresponding to 10 seconds of exposure time for each image and a total of 1 hour of observation. We also assume 60\,nm\,RMS of static aberrations and 6\,nm\,RMS of quasi-static aberrations. The zenithal angle of the star at the meridian is 20 degrees for a total field rotation of 50 degrees, with atmospheric dispersion and considering several filters. The simulations are performed with the small focal mask (radius: 25.34\,mas), which is the most critical and the Lyot stop defined previously. Coronagraphic images are post-processed with a classical Angular Differential Imaging (cADI) method, to remove the stellar residuals. Speckles are calibrated with the median of the cube of coronagraphic images and subtracted to each image. Afterwards, the calibrated coronagraphic images are aligned in the same direction and stacked.

After that, the ADI image is converted into photon flux, including photon noise, electronic noise, background, quantum efficiency of the MICADO's camera, atmospheric and telescope transmission. For each filters we determine the detection limits at $5\sigma$, normalized by the maximum of the PSF, as a function of the angular separation. These detection limits are computed by the standard deviation in an annulus for each angular separation.

A set of fake planets is simulate and compared to the detection limits \cite{2012RSPTA.370.2765A}. In this paper we present the results obtained for a 10\,Mys-old M0V-type star at 10\,pc, which corresponds to the debris disk host star, AU Mic. Fake planets have temperatures of 700, 900 and 1200\,K, which corresponds to 2, 3 and 5 Jupiter mass, respectively with a radius of one Jupiter. The projected separations are 50, 100, 150, 200 and 500\,mas, which corresponds to 0.5, 1, 1.5, 2 and 5\,au in this case.

An issue of the cADI post-processing is the self-subtraction of companions, which affects the accuracy of the photometry and the astrometry. To take care of this bias we calibrate the self-subtraction effect for each separation. As detection limits are directly compared to the flux of the fake planets, the self-subtraction effect is added to the flux of the fake planet. The same process is applied to the focal plane transmission, which is not equal to 100\% for separation closer than 2$\lambda$/D after the mask radius. 

\begin{figure}[!h]
\centering
\includegraphics[width=0.49\textwidth,clip]{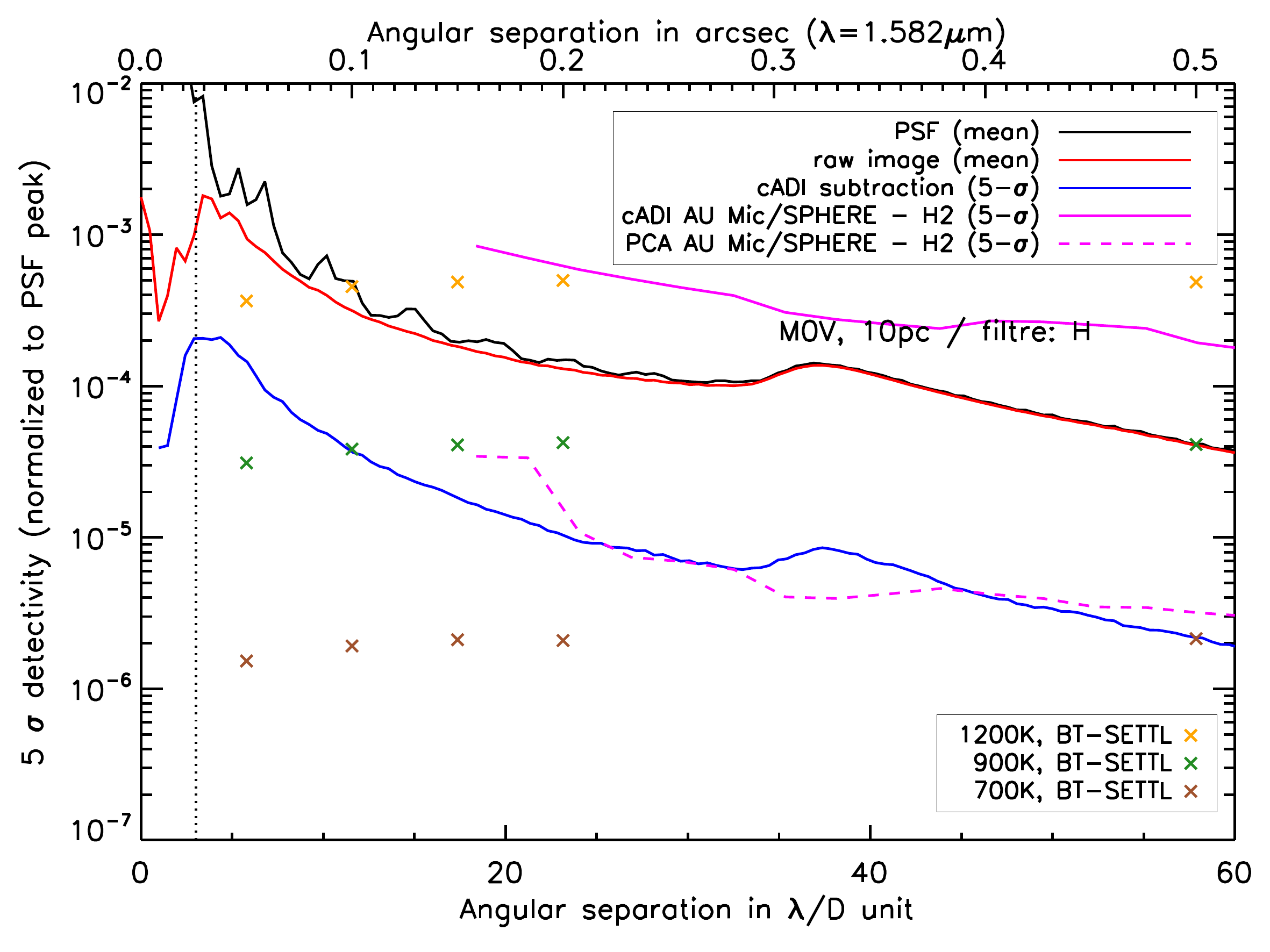}
\includegraphics[width=0.49\textwidth,clip]{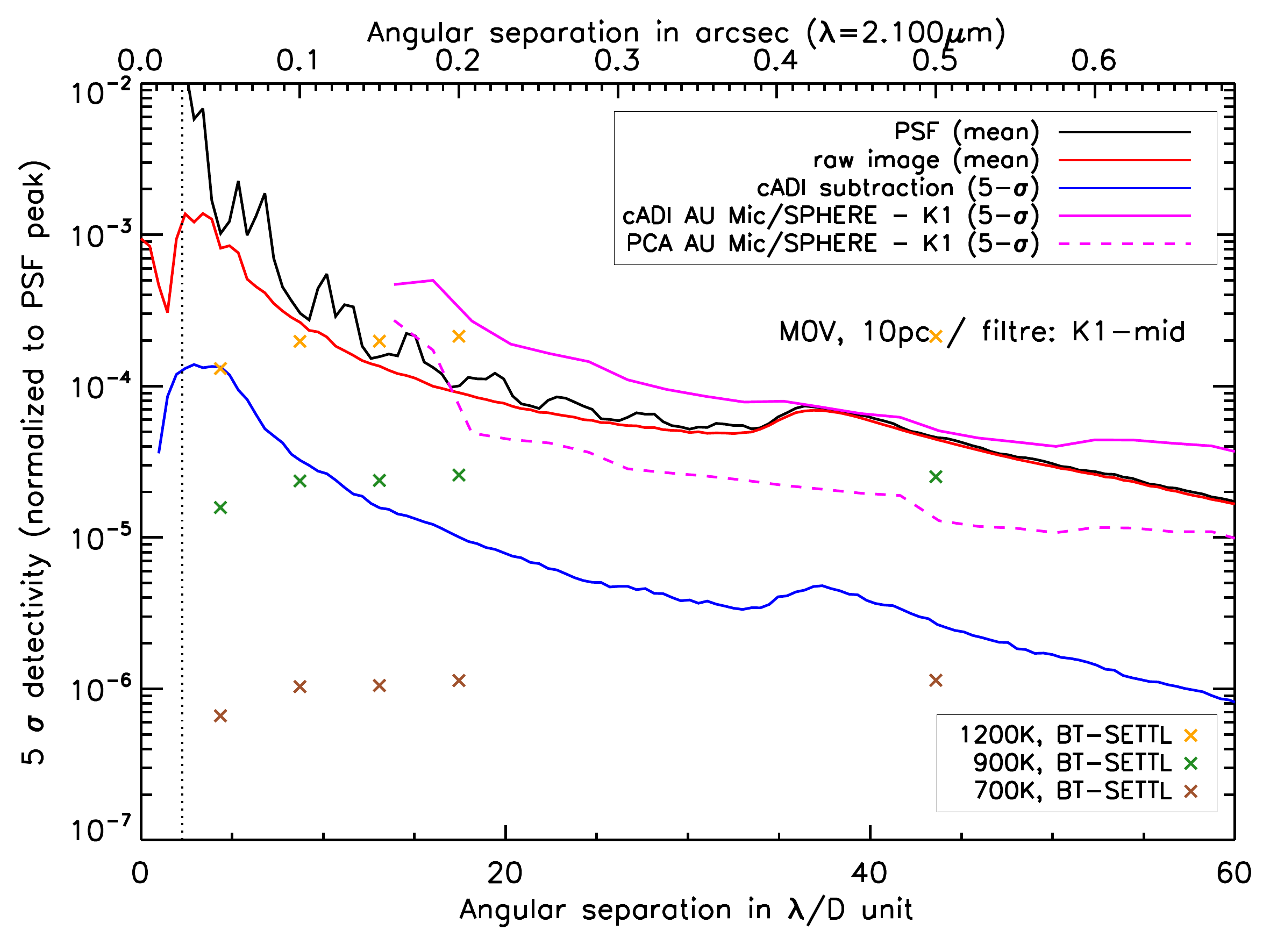}
\caption{Left: detection limits at 5\,$\sigma$ for 1 hour ADI simulation in narrowband H (CH4-short) for a 10 My M0V star at 10\,pc and on-sky observations in narrowband H2 with SPHERE of AU Mic (10 My M0V star at 10\,pc) as a function of the angular separation in arcsec (top graduation). Bottom graduation is the angular separation in $\lambda$/D for the MICADO simulation only. Blue is the detection limit of MICADO ADI simulation, magenta is detection limit of SPHERE observation with cADI (full line) and PCA (dash line) reduction, black is the radial profile of the PSF, red is the radial profile of a raw image of MICADO simulation and the cross are the signal of simulated exoplanet with temperature of 1200\,K (orange), 900\,K (green) and 700\,K (brown). Right: same as left plot but for K1-mid filter for MICADO and K1 filter for SPHERE observation of AU Mic.}
\label{detlim}
\end{figure}

Finally, the detection limits and the flux of fake planets are compared to the detection limits of observations of AU Mic with SPHERE, with similar filters. Figure \ref{detlim} shows the results for filters H2 (SPHERE) and CH4-short (MICADO) at 1.582\,$\mu$m (left) and for filters K1 (SPHERE) and K1-mid (MICADO) at 2.1\,$\mu$m (right). MICADO detection limits are in blue and the detection limits of SPHERE are in magenta for cADI (line) and PCA (Principal Component Analysis, dash). The Cross symbols correspond to the flux of the simulated fake planets. Two clear important points are revealed from these detection limits: 1) with MICADO we can access to separations which are not available for SPHERE, with a similar contrast. This range of new separation ranges from $\sim$50\,mas to 150\,mas (the current SPHERE limit). 2) The detection limits in H band is very similar, but in K band, the MICADO detection limit is better than the SPHERE detection limit. This gap is mainly due to the strong SPHERE instrumental background in K band, but it should also be noted that the instrumental background of MICADO is currently deduced from theoretical studies. 

\section{Conclusion}
\label{sec:conclu}

The preliminary design of the CLC of the high contrast imaging mode for MICADO will provide comparable contrast than SPHERE at the same angular separation, with a gain between 50 and 150\,mas, which are inaccessible for SPHERE. The CLC is composed by two focal plane masks, one small to take advantage of the angular resolution of the ELT and a medium one to minimize the effect of the atmospheric dispersion. These two masks have a radius of 25 and 50\,mas respectively and the Lyot stop, which is unique for the two focal plane masks, has a diameter of 88\% of the pupil diameter and a central obscuration of 40\%, for a total of 66,1\% of transmission.
Simulations of observations show that using broadband filters degrade the image quality, due to the atmospheric dispersion. In these conditions narrowband filters are recommended to avoid this effect. These simulations shows that for MICADO, the contrast is dominated by the AO residual aberrations and consequently the effect of the static and quasi-static aberrations are not significant with respect to SPHERE.
ADI simulations show that a gain in angular separation is possible with respect to GPI and SPHERE, with possible detection of hot and massive ($>$\,700\,K \& $>$\,2\,$M_J$) exoplanets between 50 and 150\,mas. Moreover, a gain of contrast is expected in K band, because SPHERE is not optimized for this band. However, these simulations are preliminary and some perturbations have not been taken into account yet, like jitter effect of the PSF or the real instrumental background of the entire instrument (MICADO, MAORY and the ELT). But, despite these hypotheses, simulations are still instructive and give the potential of the high contrast imaging mode of MICADO.
The next step of the study is to merge the code of high contrast imaging and the COMPASS program. This merging will provide fast simulations and the possibility to add different features like jitter, Vortex coronagraph, etc.

\acknowledgments 
 
The authors thank the MICADO team for its contribution to the design of the instrument. The authors thank ESO, CNRS/INSU and Observatoire de Paris for their financial support. 
\bibliography{report} 
\bibliographystyle{spiebib} 

\end{document}